\documentclass[10pt,letterpaper]{article}
\usepackage[top=0.85in,left=1.5in,footskip=0.75in,marginparwidth=2in]{geometry}

\usepackage[utf8]{inputenc}

\usepackage{cite}

\usepackage{nameref,hyperref}

\usepackage{microtype}
\DisableLigatures[f]{encoding = *, family = * }

\usepackage{changepage}

\usepackage[aboveskip=1pt,labelfont=bf,labelsep=period,singlelinecheck=off]{caption}

\makeatletter
\renewcommand{\@biblabel}[1]{\quad#1.}
\makeatother

\usepackage{lastpage,fancyhdr,graphicx}
\usepackage{color}

\definecolor{Gray}{gray}{.25}

\usepackage{graphicx}

\usepackage{sidecap}

\usepackage{wrapfig}
\usepackage[pscoord]{eso-pic}
\usepackage[fulladjust]{marginnote}
\reversemarginpar

\begin{document}
\vspace*{0.35in}

\begin{flushleft}
{\Large
\textbf\newline{Accurate laser frequency locking to optical frequency combs under low-signal-to-noise-ratio conditions}
}
\newline
\\
C. Guo\textsuperscript{1},
M. Favier\textsuperscript{1},
N. Galland\textsuperscript{1},
V. Cambier\textsuperscript{1},
H. \'Alvarez-Mart\'inez\textsuperscript{1,2},
M. Lours\textsuperscript{1},
L. De Sarlo\textsuperscript{1}
M. Andia\textsuperscript{1}
R. Le Targat\textsuperscript{1}
S. Bize\textsuperscript{1,*}
\\
\bigskip
\bf{1} LNE-SYRTE, Observatoire de Paris, Universit\'e PSL, CNRS, Sorbonne Universit\'e, 61 avenue de l'Observatoire, 75014 Paris, France
\\
\bf{2} Real Instituto y Observatorio de la Armada, Calle Cecilio Pujaz\'on, San Fernando, 11110 C\'adiz, Spain
\\
\bigskip
* sebastien.bize@obspm.fr

\end{flushleft}

\section*{Abstract}
We demonstrate a method for accurately locking the frequency of a continuous-wave laser to an optical frequency comb in conditions where the signal-to-noise ratio is low, too low to accommodate other methods. Our method is typically orders of magnitude more accurate than conventional wavemeters and can considerably extend the usable wavelength range of a given optical frequency comb. We illustrate our method by applying it to the frequency control of a dipole lattice trap for an optical lattice clock, a representative case where our method provides significantly better accuracy than other methods.



\section{\label{sec_Introduction}Introduction}

Self-referenced optical frequency combs provide an accurate optical frequency reference over a wide optical bandwidth \cite{Hall2006, Hansch2006, Diddams2010}. They can be used to measure or stabilize the frequency of continuous-wave (CW) lasers in many applications such as state-of-the-art optical-frequency metrology \cite{Ludlow2015, Bize2019}, precision spectroscopy \cite{Holzwarth2000}, laser cooling of atoms \cite{Fordell2014, Mcferran2018}, exciting rovibrational transitions in cold molecules \cite{Prehn2017}, measurement of the frequency of stabilized lasers for length metrology \cite{Hong2016}, etc. This relies on exploiting the beat note between the CW laser and a tooth of the comb, either by frequency counting or by phase locking\cite{Holzwarth2000, Jones2000, Cundiff2001}. These methods can provide exquisitely high accuracy but require a beat note with high signal-to-noise ratio. Too low a signal-to-noise ratio will lead to cycle slips and improper counting or phase locking \cite{Oscillator, Ascheid1982, Gerberding2013, Francis2014}. This limits the range of conditions over which these methods apply. For instance, the wavelength range is limited to spectral regions where the power per tooth of the comb is significant. Similarly, these methods are limited to conditions where the CW laser power available for the beat note is high enough and where the laser linewidth is narrow enough. In this article, we demonstrate a method to accurately lock a laser to an optical frequency comb in conditions of low signal-to-noise ratio and/or broad linewidth (hundreds of kHz or more). While being nominally less accurate than counting or phase-locking, our method is orders of magnitude more accurate than conventional wavemeters \cite{Kobtsev2007, Saleh2015} and can considerably extend the usable wavelength range of a given optical frequency comb. Also, it is more accurate than other simple schemes used to stabilize lasers to optical frequency combs in conditions of low signal-to-noise ratio \cite{Fordell2014, Prehn2017}. Our method is applied to controlling the frequency of a lattice trap in an optical lattice clock \cite{Tyumenev2016}, which is a representative case where our method is an efficient choice.

The article is arranged as follows. The locking architecture will be introduced in section II by considering frequency-locking of a single-frequency CW laser to an optical frequency comb. In section III, we describe the synthesis of the 362.5~nm lattice trap light of our optical lattice clock setup, and we demonstrate how the proposed locking scheme is applied to stabilizing its frequency. Finally, in order to test the accuracy of our locking scheme, a voltage-controlled oscillator is used in section IV as a laser beat note emulator for frequency counting. The results show that our locking scheme has an accuracy of less than 1~kHz in our specific experimental conditions. In the last section, a beat note with extremely low signal-to-noise ratio is locked to demonstrate the robustness of the locking scheme to signals with low signal-to-noise ratio.

\section{\label{sec_LockingScheme}Locking system architecture}

\begin{figure}[ht]
    \centering
    \includegraphics[width=0.75\columnwidth]{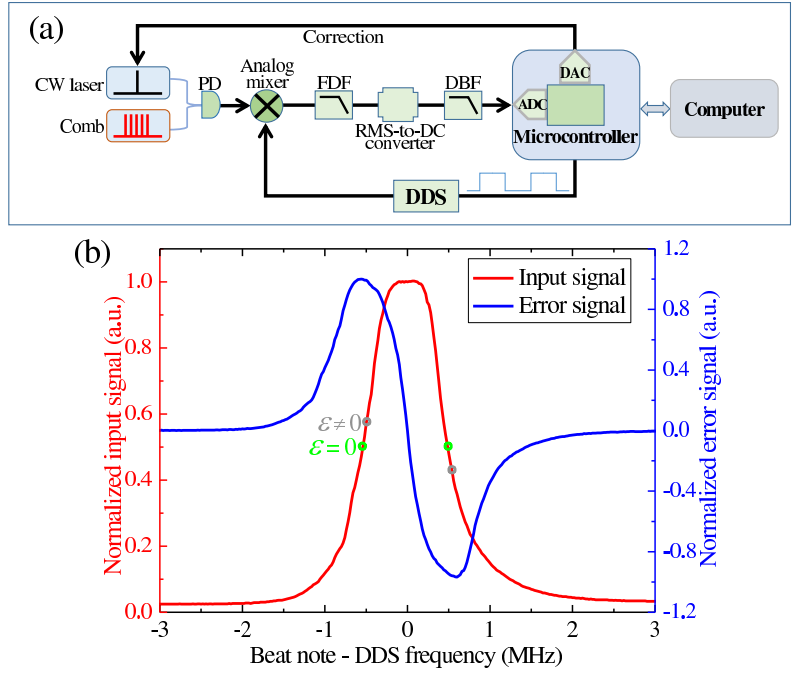}
    \caption{(a) Schematic of the locking loop. FDF: frequency discriminating filter; DBF: detection bandwidth filter; RMS: root-mean-square; DC: direct current; ADC/DAC: analog-to-digital/digital-to-analog converter. The rectangular wave before DDS represents the modulation of the DDS frequency by the microcontroller, which can also communicate with a computer for collecting data and receiving commands. Other optical and electronics components used for the implementation of the scheme \cite{Infor2019} are not shown here for simplification. (b) Typical normalized input signal (output of the beat note detection chain digitized by the ADC) (red solid line) and error signal (blue solid line) as the frequency offset of the DDS is scanned. The green and gray circles on the transmission curve represent two different cases that lead to an error signal $\epsilon = 0$ and $\epsilon \not= 0$, respectively.}
\label{Fig_1_single_loop_locking_scheme_new.eps}
\end{figure}

In the following, we consider the beat note between a CW laser and a comb tooth from an optical frequency comb to illustrate the principle of our locking scheme. As shown in Figure \ref{Fig_1_single_loop_locking_scheme_new.eps}(a), the frequency of the beat note is mixed down to a near-zero frequency with a microcontroller-controlled direct-digital-synthesizer (DDS) signal. The frequency of the DDS is alternated between two symmetrical values around the target locking frequency (square-wave frequency modulation). The near-zero frequency produced by the analog mixer then passes through a frequency-discriminating filter (FDF), which is a low-pass filter. A root-mean-square-to-direct-current (RMS-to-DC) converter transforms the filtered signal into a DC signal, before filtering by the detection-bandwidth filter (DBF), which is also a low-pass filter with lower bandwidth than that of FDF. After passing through the analog-to-digital converter (ADC), the signal is further processed by the microcontroller synchroneously with the frequency modulation. The microcontroller then implements a digital lock-in demodulation to compute the error signal. This error signal is further processed through a digital proportional-integrator (PI) loop filter within that same microcontroller. A digital-to-analog converter (DAC) finally allows closing the feedback loop by generating an analog correction controlling the PZT actuator of the laser. The FDF enables frequency discrimination by producing an effective symmetrical peak-like feature around the target locking frequency (see Figure \ref{Fig_1_single_loop_locking_scheme_new.eps}(b)). The modulation/demodulation scheme allows the beat note to be locked to the mean of the two DDS frequencies, therefore providing an accurate lock. The DBF plays the role of an anti-aliasing filter for the subsequent digitization process.

A typical normalized input signal (i.e. the output of the beat note detection chain digitized by the ADC) and the associated error signal are represented in Figure \ref{Fig_1_single_loop_locking_scheme_new.eps}(b) as the frequency offset of the DDS is scanned. When the loop is closed, we use the recorded error signal to estimate the frequency stability and accuracy of the laser, while the correction signal is used to estimate its free-drifting behavior. Optimal locking conditions are reached by properly setting the gain in the loop, the bandwidths of the FDF and DBF, the modulation depth of the DDS, as well as the parameters in the digital PI loop. In order to adapt to different locking conditions, the FDF and DBF bandwidths, as well as the DDS modulation period, should be tuned accordingly: the first one determines the quality of frequency discrimination (and needs to be adapted to the laser linewidth), while the second and third ones are interrelated with the iteration rate of the loop.

\section{\label{sec_ApplicationToClock}Application to an optical lattice clock}

Optical lattice clocks have achieved unprecedented levels of performance \cite{Mcgrew2018, Oelker2019} with fractional frequency instability approaching $1\times 10^{-18}$. This has greatly benefited from the ``magic wavelength'' (or ``magic frequency'') scheme \cite{Katori2003}, which allows the trapping of the atomic ensemble with unperturbed transition frequency while maintaining the other merits of the system. Proper implementation of this scheme requires sufficiently stable, accurate and agile control of the lattice trap frequency. As an example, in our mercury optical lattice clock \cite{Tyumenev2016}, an error of 1~MHz on that frequency leads to a significant fractional clock frequency shift of $-1.1\times10^{-17}$ at a lattice depth of 100 $E\rm_{rec}$ (recoil energy). In this section, we demonstrate that the proposed frequency-locking scheme can meet these stability, accuracy and agility requirements.

\begin{figure}[ht]
    \centering
    \includegraphics[width=0.75\columnwidth]{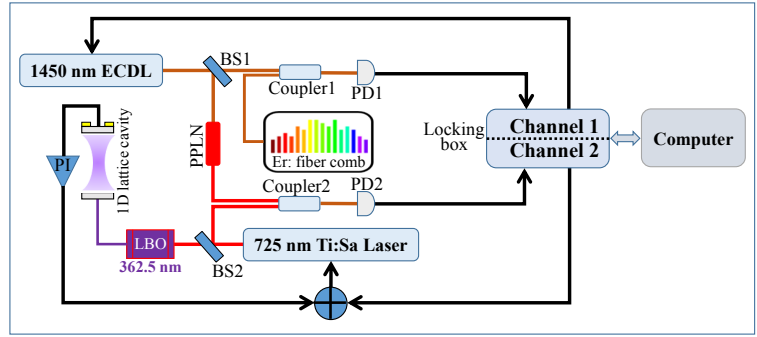}
    \caption{Application of the proposed locking scheme to control the frequency of a lattice trap for an optical lattice clock. ECDL: external cavity diode laser; BS: beam splitter; PPLN: periodically poled lithium niobate; Ti:sa: Titanium:sapphire laser; LBO: lithium triborate. The locking box is comprised of two feedback loops working on the same principle, as described in section II. The first loop (Channel 1) is used to lock the 1450~nm ECDL to an optical frequency comb (Er:fiber comb). Channel 2 is then used to lock the 725~nm Ti:sa laser to the frequency-doubled ECDL. In order to reach the magic wavelength (close to 362.5~nm), the Ti:sa laser is finally frequency-doubled using the LBO crystal. The 1D lattice cavity is used to enhance the lattice laser intensity available for the atomic transition interrogation \cite{Tyumenev2016}.}
\label{Fig_2_two_loop_locking_scheme.eps}
\end{figure}

The magic wavelength for the isotope $^{199}$Hg is about 362.5~nm. Such a wavelength is obtained from frequency-doubling the light from a 725~nm Ti:sa laser through a lithium triborate (LBO) crystal (see Figure \ref{Fig_2_two_loop_locking_scheme.eps}). In order to precisely control the laser frequency, we would like to lock our Ti:sa laser directly to an optical frequency comb. However, due to the lack of spectral power at 725~nm of our Er:fiber-based comb, an intermediate laser working at 1450~nm is used for frequency transfer. The 1450~nm external cavity diode laser (ECDL) is locked to an Er:fiber optical frequency comb with the proposed locking scheme (as described in section \uppercase\expandafter{\romannumeral2}), using the beat note between a fraction of the ECDL light and a comb tooth. The rest of the ECDL light is frequency-doubled to 725~nm through a periodically poled lithium niobate (PPLN) crystal, and is then locked to the Ti:sa laser thanks to a second frequency-locking loop with the same working principle. Additionally, a digital feed-forward loop acting on the PZT voltage of the Ti:sa laser was implemented to allow for dynamical laser frequency (wavelength) tuning (not shown in Figure \ref{Fig_2_two_loop_locking_scheme.eps}). This feature is notably required to realize differential measurements between configurations with different optical lattice wavelengths, which has crucial applications such as precisely determining the lattice magic wavelength for our atomic system \cite{Tyumenev2016}.

\begin{figure}[ht]
    \centering
    \includegraphics[width=0.75\columnwidth]{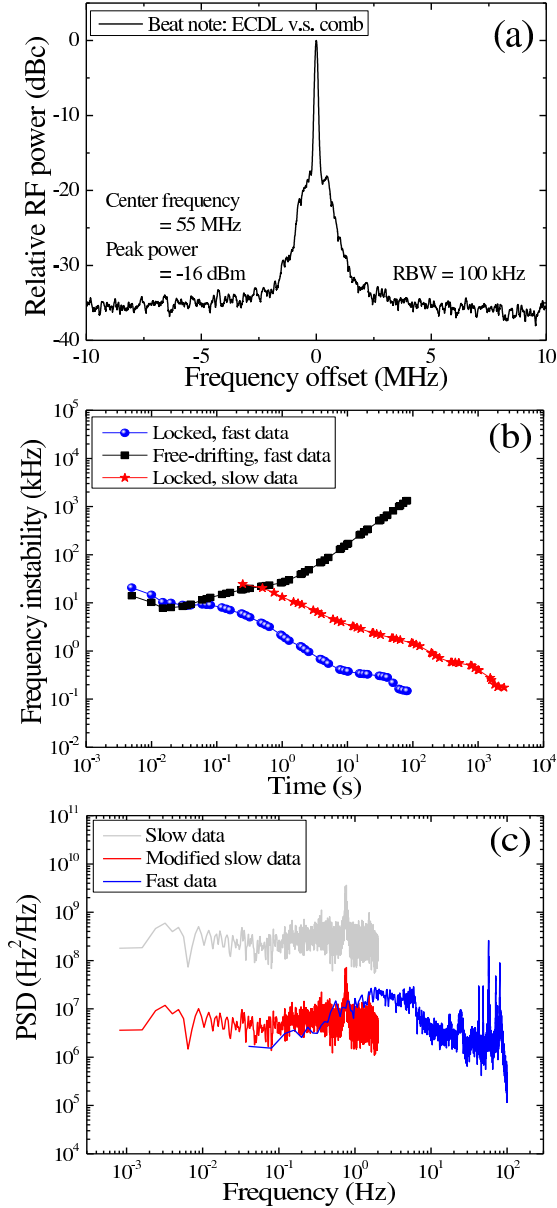}
    \caption{(a) A typical beat note between the ECDL and the comb, measured with a resolution bandwidth (RBW) of 100~kHz. The RF power is normalized to its peak power. (b) Frequency instabilities of the ECDL laser calculated from fast data and slow data, respectively. The fast data have a time interval of 5~ms (same as the iteration period $T=5$~ms), while the slow data are extracted from the fast data with a time interval of 250~ms. The locked traces use the error signal as the data source while the free-drifting trace uses the correction signal. All of the frequency instabilities are based on Allan standard deviations in this work except stated otherwise. (c) Power spectral density (PSD) corresponding to (b): the red curve is obtained from the gray curve by dividing by a factor of 50 to account for the aliasing.}
\label{Fig_3_onlyECDL_new.eps}
\end{figure}

The performance of the first locking loop (for locking the ECDL to the comb) is shown in Figure \ref{Fig_3_onlyECDL_new.eps} for demonstration. Figure \ref{Fig_3_onlyECDL_new.eps}(a) is a typical beat note between the ECDL and one comb tooth, where the signal is normalized to the peak power. The beat note has a signal-to-noise-ratio of $\sim$35~dB in a resolution bandwidth (RBW) of 100~kHz. Figure \ref{Fig_3_onlyECDL_new.eps}(b) and \ref{Fig_3_onlyECDL_new.eps}(c) show the frequency instability based on Allan standard deviation (ADEV) and the power spectral density (PSD), respectively. Here we used bandwidths of $\sim$500~kHz for the FDF and $\sim$5~kHz for the DBF, a modulation depth of the DDS of 500~kHz, and an iteration period of $T = 5$ ms. The ``fast data'' are obtained by collecting data from the microcontroller at the end of each iteration, i.e.\ every 5~ms. The ``slow data'' are obtained by only extracting data once every 50 iterations. Information extracted from the fast and slow datasets allows us to estimate both the short-term (high-frequency) and long-term (low-frequency) stabilities. In contrast to the locked traces which use the error signal $\epsilon$ as the data source, the free-drifting trace uses the correction signal. As shown in Figure \ref{Fig_3_onlyECDL_new.eps}(b), once the loops are closed, the frequency instabilities evolve nearly as $\tau^{-1/2}$ (where $\tau$ is the integration time) in both traces (blue solid dots and red triangles), indicating that white frequency noise dominates. The free-drifting trace (black squares) decreases at first and then continuously increases as $\tau^{+1/2}$, showing a strong character of random-walk noise. Both frequency instabilities after a single iteration (5 ms) are smaller than 30~kHz in the locked traces. We also estimated the frequency instability of the second loop, which is typically smaller than 100~kHz after a single iteration, as will be shown in Figure \ref{Fig_7_LSNRtest_new.eps}. Taking the frequency doubling (see Figure \ref{Fig_2_two_loop_locking_scheme.eps}) into account, frequency fluctuations of the lattice light translate into instability of the clock frequency below 5 parts in $10^{18}$ (at worst) after a single iteration, which satisfies our requirement. The two locked traces in Figure \ref{Fig_3_onlyECDL_new.eps}(b) do not overlap because the slow data's sampling rate is 50 times as low as the fast data's. Therefore, they show nearly the same instability at their first points and nearly the same evolution with time when the noise is dominated by the white frequency noise. This white-frequency-noise-like behavior is also clearly shown in the power spectral density (PSD) estimated from the same data and displayed in Figure \ref{Fig_3_onlyECDL_new.eps}(c).

\section{\label{sec_Accuracy}Accuracy of the frequency lock }

Our scheme is aimed at locking a laser to a reference laser (such as an optical frequency comb) under conditions of low signal-to-noise ratio and/or broad linewidth. This provides a basis for an accurate control of its frequency, as long as the reference laser itself is connected to a primary frequency standard. It is therefore relevant to assess the contribution of our scheme to the accuracy of the overall system.

\begin{figure}[ht]
    \centering
    \includegraphics[width=0.75\columnwidth]{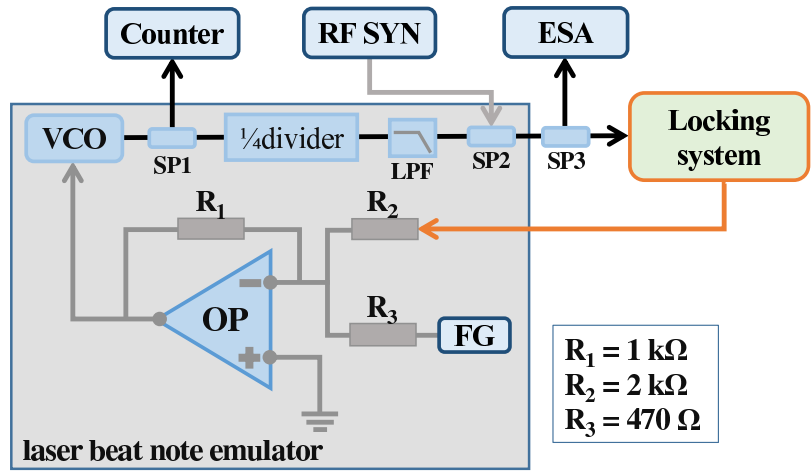}
    \caption{Schematic of laser beat note emulator for accuracy tests. A voltage-controlled oscillator (VCO) is used to simulate the laser beat note. An operational amplifier is used to add the correction signal from the locking system and the noise signal (the offset is tuned to reach a target frequency of 45~MHz) from a function generator. Three power splitters are used for frequency counting (SP1), asymmetric signal adding (SP2) and frequency monitoring (SP3), respectively. LPF: low pass filter; ESA: electronic spectrum analyser; RF SYN: RF synthesizer; FG: function generator. A 1/4 divider is used for frequency division to a frequently-used range in the locking system. The resistances shown are selected to balance the gain. Note that the noise and the sidebands are not injected at the same time.}
\label{Fig_4_VCOtest.eps}
\end{figure}

In our scheme the lock accuracy is principally granted by the way we make use of the FDF filter and the frequency-modulation of the DDS. In this section, we describe further tests made to quantify this accuracy. To this end, we designed a laser beat note emulator as shown in Figure \ref{Fig_4_VCOtest.eps}. A voltage-controlled oscillator (VCO) simulates the laser beat note. An operational amplifier implements the correction signal from the locking system, and can also add noise from an external noise source. Three power splitters are used for frequency counting (SP1), asymmetric sideband signal injecting (SP2) and frequency monitoring (SP3), respectively. Because our Ti:sa laser and ECDL have typical linewidths of around 100~kHz, the noise bandwidth is fixed at $f\rm_{BW}$ = 100~kHz. The sideband signal is added to simulate asymmetric signals as will be shown later. A 1/4 frequency divider is used for frequency division to reach a low frequency range before reaching the locking system. 

\begin{figure}[ht]
    \centering
    \includegraphics[width=0.5\columnwidth]{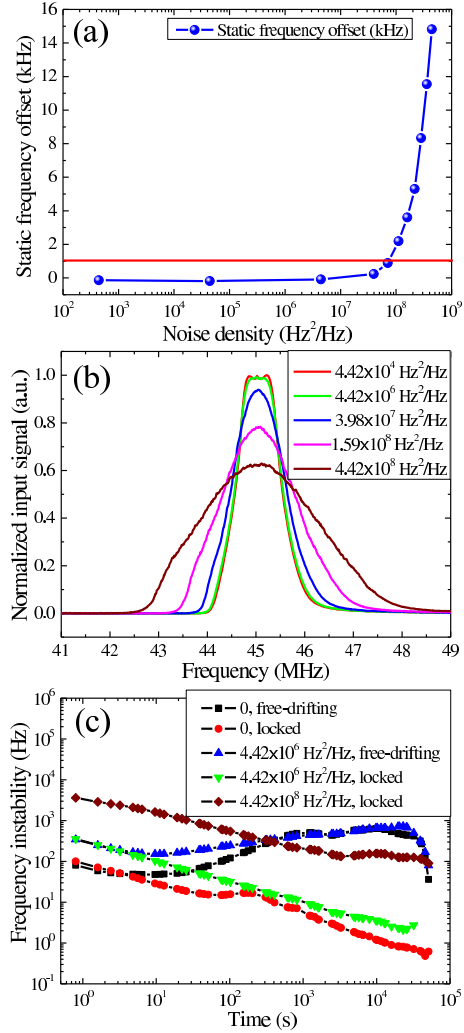}
    \caption{Accuracy tests of the laser beat note emulator with broad linewidth. (a) Static frequency offset of laser beat note emulator versus noise density (see text). The red solid line indicates a frequency offset of 1~kHz. (b) Normalized input signals when scanning the DDS frequency offset (similar to Figure 1(b)) in conditions of different noise densities. The red and green curves are dominated by the profile of the FDF, while the blue, pink and wine curves are dominated by the profiles of the noise (corresponding to strongly degraded operation conditions). All the curves are normalized to peak input signal in nominal conditions (maximum of the red curve). (c) Frequency instabilities of the emulator with different noise densities (red, green and wine curves). Zero means no noise is injected. The free-drifting traces (black and blue curves) under conditions of different noise densities are also added for comparison.}
\label{Fig_5_lock_VCO_newfigure.eps}
\end{figure}

We first use the emulator to simulate a broad laser beat note by injecting noise to it with a function generator (FG). Note that no sideband signals are injected through SP2 in this test. In the following, we will characterize the injected noise by its noise density, $S\rm_{\nu}$, defined as ($K$(Hz/V)$\cdot$ $A\rm_{RMS}$(V)/4)$^2$/$f\rm_{BW}$(Hz), where $K$ is the frequency tuning coefficient of the VCO, $A\rm_{RMS}$ is the effective voltage of the noise signal (the gain in the operational amplifier is considered here), and the factor of 4 accounts for the presence of the divider in the laser emulator (see Figure \ref{Fig_4_VCOtest.eps}). The injected noise is therefore a white frequency noise up to the cutoff frequency $f\rm_{BW}$. The test results are summarized in Figure \ref{Fig_5_lock_VCO_newfigure.eps}. Figure \ref{Fig_5_lock_VCO_newfigure.eps}(a) shows the static frequency offset (the difference between the measured mean frequency and the target frequency) of the emulator as a function of the noise density. As shown in Figure \ref{Fig_5_lock_VCO_newfigure.eps}(a), the static frequency offset greatly increases once the noise density is more than 10$^8$ Hz$^2$/Hz. The red solid line indicates a frequency offset of 1~kHz, which we chose as a criterion for good accuracy. This sets a critical noise density of about $7.1\times10^7$ Hz$^2$/Hz. The input signals when scanning the DDS frequency offset are plotted under different conditions of noise density, as shown in Figure \ref{Fig_5_lock_VCO_newfigure.eps}(b). The input signals show a similar profile to the FDF's profile when the noise densities are low (red and green curves), while they are dominated by the noise profile when the noise densities are high (blue, pink and wine). For instance, in the case of the blue solid line with a noise density of $3.98\times10^7$ Hz$^2$/Hz (a situation that is already much worse than our typical operation conditions), the beat note linewidth is about 1.2~MHz. This means that the noise has already exceeded the effective bandwidth of the FDF, yet the corresponding static frequency offset is only 234~Hz, showing the robustness of our locking scheme even in highly-noisy conditions. Ideally, high noise levels should not deteriorate the locking static frequency offset, provided the signal keeps a symmetric profile. However, Figure \ref{Fig_5_lock_VCO_newfigure.eps}(a) shows an increase of the locked static frequency offset with increasing noise density. A possible explanation lies in the asymmetry of the emulated beat note.

The frequency instabilities of the emulator under different noise densities are also shown in Figure \ref{Fig_5_lock_VCO_newfigure.eps}(c). From the red, blue and olive dots, there is a clear overall increase of the instability with increased noise densities. The slight and broad bump in the middle of the red curve might come from the VCO itself or from the locking system. Otherwise, all three curves show favorable averaging behavior (close to a $\tau^{-1/2}$ power law), meaning that the white frequency noise also dominates here. From the blue to olive curves in Figure \ref{Fig_5_lock_VCO_newfigure.eps}(a), the instability increases nearly by a factor of 10 when the noise is increased by a factor of 100, since the ADEV is proportional to $\sqrt{S\rm_{\nu}}$. The free-drifting curves from correction signals similar as shown in Figure \ref{Fig_3_onlyECDL_new.eps}(b) with (magenta dots) and without injected noise (black dots) are also shown for comparison.

\begin{figure}[ht]
    \centering
    \includegraphics[width=0.75\columnwidth]{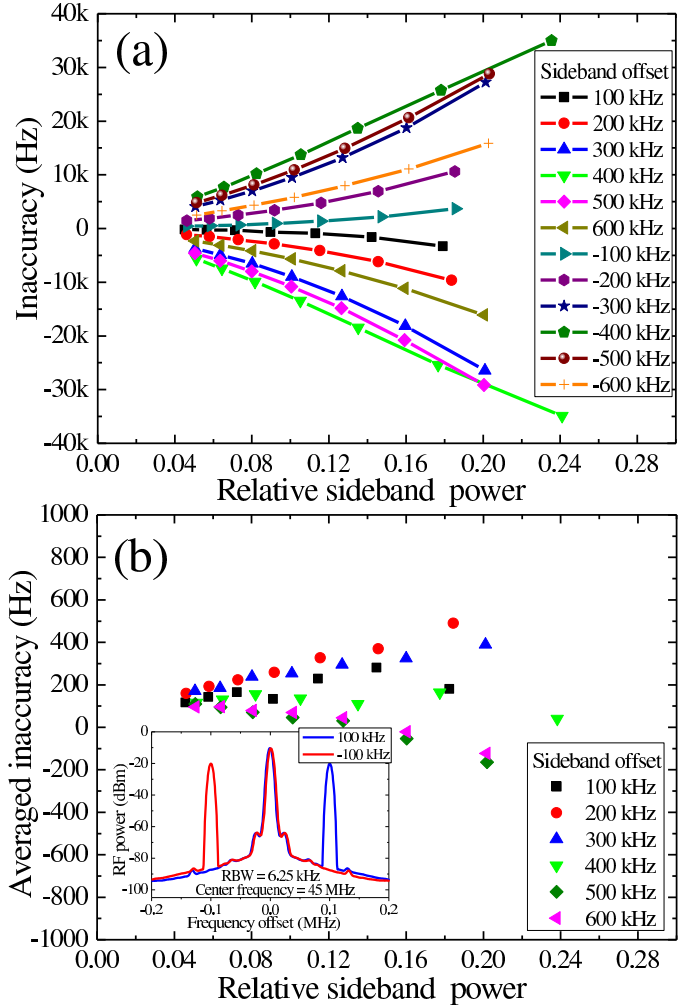}
    \caption{Accuracy tests of the laser beat note emulator with asymmetric sidebands. (a) Static frequency offsets for single positive or negative sidebands with different relative powers. Only one sideband is added to the carrier (45~MHz) for each data point. (b) Mean static frequency offsets with different sideband offset frequencies and different relative sideband powers (each point is a mean of the positive and negative sidebands at the same frequency). Mean static frequency offsets are no more than 1~kHz, meaning that in our configuration, the locking system has an accuracy better than 1~kHz. The inset in (b) shows the asymmetric beat notes with the +100~kHz and -100~kHz sidebands, respectively (RBW = 6.25~kHz).}
\label{Fig_6_lock_VCO_Asymmetry.eps}
\end{figure}

Optical beat notes often exhibit spectral features other than the component of interest, in particular when an optical frequency comb is involved. We therefore designed tests to estimate how the presence of spurious spectral features in the input beat note affects the static frequency offset. For these tests we added a single sideband (with controllable relative power and frequency detuning) to the beat note carrier, using SP2 (see Figure \ref{Fig_4_VCOtest.eps}). 
Examples of such asymmetric signals obtained with a +100~kHz or -100~kHz sideband (relative power $\sim$ 0.1) are represented in the insert of Figure \ref{Fig_6_lock_VCO_Asymmetry.eps}(b) for illustration. In Figure \ref{Fig_6_lock_VCO_Asymmetry.eps}(a), the tests reveal that the static frequency offsets are increasing with increased sideband power in a way that depends on the sideband frequency offsets. Even though the static offset frequencies greatly increase with single sidebands (representing cases of asymmetry), the mean static frequency offsets are never more than 1~kHz in our measurements, as shown in Figure \ref{Fig_6_lock_VCO_Asymmetry.eps}(b). In our application case (see Figure \ref{Fig_2_two_loop_locking_scheme.eps}), spurious spectral features which are or may be present are orders of magnitude lower than in these tests, therefore resulting in static frequency offsets much below 1~kHz.

Fundamentally, the accuracy of our method relies on the accuracy of the digital synthesis and on the square-wave frequency modulation. At the equilibrium of the servo loop, the FDF is used at the same frequency on both sides of the modulation sequence. As a consequence, details and imperfections of the filter characteristics do not lead to any bias in the frequency lock. The error on the lock point would typically be no more than a small fraction of the FDF bandwidth. This is what we reported earlier in this section. Next, we mention effects that can cause a bias because they introduce a difference in the response of the detection chain between the two sides of the modulation sequence. A first effect can come from a slope in the background of the spectrum of the beat note. For example, in our conditions, we observe a background from the optical frequency comb about 24~dB below the signal of interest and with a slope of about 1~dB over 20~MHz. This causes a bias on the order of 10~Hz. A second effect can come from spurious sidebands in the spectrum. For instance, a sideband 7~MHz away from the main beat note, with a power 21~dB below it (as may occasionally be observed in our conditions), could yield a bias of less than 10~Hz. The DDS itself has spurious features in its spectrum which are in general asymmetric with respect to the main output frequency. The corresponding bias is typically small in our conditions.

It is worth noting that, in the view of other applications of the scheme, the sensitivity to spectral perturbations can be optimized, in particular by adapting the bandwidth of the FDF. For example, for measuring or locking the frequency of a laser with a 10~kHz linewidth, one could use an FDF with a bandwidth of $<50$~kHz and reduce the sensitivity to spectral perturbations by more than one order of magnitude compared to our settings. 

\section{\label{sec_LockingWithNoise}Locking with low signal-to-noise ratio}

\begin{figure}[ht]
    \centering
    \includegraphics[width=0.75\columnwidth]{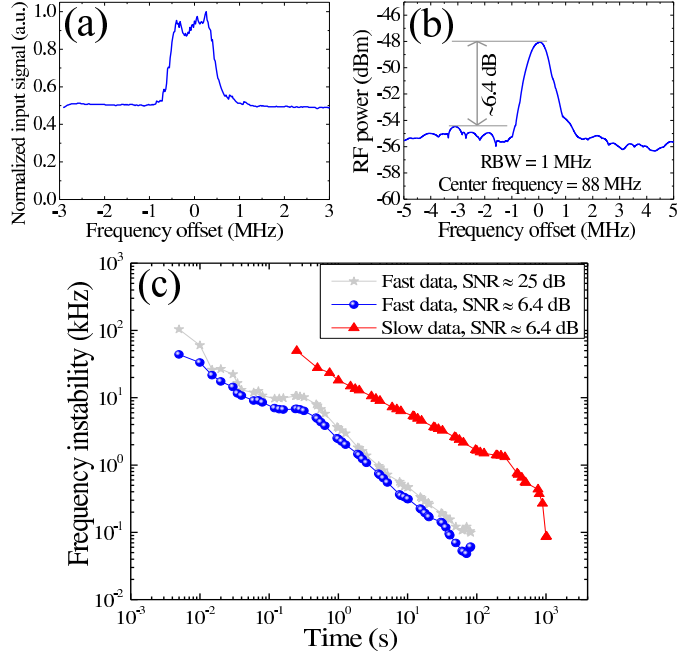}
    \caption{Lock accuracy test in low-signal-to-noise-ratio conditions. (a) Normalized input signal when scanning the DDS frequency offset under a very low-signal-to-noise-ratio condition. (b) Beat note spectrum between the Ti:sa laser and the frequency-doubled ECDL corresponding to (a). The signal is recorded with an ESA averaging 10 times with a RBW of 1~MHz. (c) Frequency instabilities for the low-signal-to-noise-ratio ($\approx$ 6.4~dB in 1~MHz RBW) condition with fast data (blue solid dots), and slow data (red triangles), respectively. A frequency instability curve (gray stars) with relatively high signal-to-noise ratio ($\approx$ 25~dB in 1~MHz RBW) is also shown for comparison.}
\label{Fig_7_LSNRtest_new.eps}
\end{figure}

In this section, we demonstrate that our method can be extended to low-signal-to-noise-ratio conditions. To this end, we will use the second servo loop (see Channel 2 in Figure \ref{Fig_2_two_loop_locking_scheme.eps}), where the beat note between the Ti:sa laser and the frequency-doubled ECDL is first attenuated greatly and then re-amplified with several amplifiers. The normalized input signal when scanning the DDS frequency offset, as well as the real beat note signal, are shown in Figure \ref{Fig_7_LSNRtest_new.eps}(a) and \ref{Fig_7_LSNRtest_new.eps}(b), respectively. From Figure \ref{Fig_7_LSNRtest_new.eps}(a), we find an input signal sitting on a non-zero background corresponding to the noise. The background amplitude is half that of the input signal peak, meaning that the signal-to-noise ratio has been strongly decreased. This is further confirmed by Figure \ref{Fig_7_LSNRtest_new.eps}(b), where the locked signal has a signal-to-noise ratio of 6.4~dB in 1~MHz RBW. The frequency instabilities for the low-signal-to-noise-ratio locking are shown in Figure \ref{Fig_7_LSNRtest_new.eps}(c). Both the fast (blue solid dots) and slow data (red triangles) average down as expected. A frequency instability curve of the same locking loop with relatively high signal-to-noise ratio (25~dB in 1~MHz RBW) is also shown for comparison. The similar instabilities between the low-signal-to-noise-ratio locking and high-signal-to-noise-ratio locking mean that our locking scheme is not sensitive to the signal-to-noise ratio of the beat notes. The slight difference between the two curves (red and gray) is mainly due to the differences between the intensities of the locked signals and gains selected in the loop. Note that the hardware was not modified for adapting to the low-signal-to-noise-ratio cases.

\section{\label{sec_Conclusions}Conclusion }

We described a method to accurately lock the frequency of a laser to a beat note with another laser in situations of broad laser linewidth and poor signal-to-noise ratio. We have tested and quantified effects that can influence the lock accuracy. Moreover, we tested the effectiveness of the method in the case of strongly degraded signal-to-noise ratio. The method is especially suited to controlling accurately the frequency of a laser with an optical frequency comb. The use of this method extends the usable wavelength range of optical frequency combs to cases where coherent control is not possible (or necessary). It enables accurate frequency control of broad-linewidth lasers and of lasers with very low power. For certain applications, it can considerably extend the practical use of a given optical frequency comb to regions of the spectrum where the optical power per tooth of the comb is low. We described a particular application of our method to the accurate frequency control of the lattice laser in an optical lattice clock. Our specific implementation of the scheme yields a control of the lattice laser frequency to the $\sim 1$~kHz level ($\sim 10^{-12}$), a value itself specific to our case. When required, key parameters of the scheme can be adapted to reach even better accuracy (if allowed by the laser's characteristics). This method can also apply to other cases such as stabilizing lasers for cooling atoms and ions \cite{Fordell2014, Mcferran2018}, length measurement \cite{Hong2016}, laser ranging \cite{Luo2015}, etc.

\section*{Acknowledgements}
	We acknowledge contributions from the SYRTE electronic workshop. This work was supported by ERC Consolidator Grant AdOC (ERC-2013-CoG-617553).
\bibliography{library}

\begin{thebibliography}{10}

\bibitem{Oscillator}
{\em Another well-known approach to deal with low-signal-to-noise-ratio optical
  beat notes is to use a tracking oscillator. As this approach relies in the
  first place on phase locking, our statements also apply to tracking
  oscillators}.

\bibitem{Infor2019}
{\em DDS: Analog Devices, AD9912A; microcontroller: Arduino YUN (ADC included);
  DAC: Analog Devices, AD5734R; RMS-to-DC converter: Analog Devices, AD536A}.

\bibitem{Ascheid1982}
G.~Ascheid and H.~Meyr.
\newblock Cycle slips in phase-locked loops: A tutorial survey.
\newblock {\em IEEE Trans. Commun.}, 30(10):2228--2241, 1982.

\bibitem{Bize2019}
S.~Bize.
\newblock The unit of time: Present and future directions.
\newblock {\em Comptes Rendus Physique}, 20(1-2):153--168, jan 2019.

\bibitem{Cundiff2001}
S.~T. Cundiff, J.~Ye, and J.~L. Hall.
\newblock Optical frequency synthesis based on mode-locked lasers.
\newblock {\em Rev. Sci. Instrum.}, 72(10):3749--3771, 2001.

\bibitem{Diddams2010}
S.~A. Diddams.
\newblock The evolving optical frequency comb.
\newblock {\em J. Opt. Soc. Am. B}, 27(11):B51--B62, 2010.

\bibitem{Fordell2014}
T.~Fordell, A.~E. Wallin, T.~Lindvall, M.~Vainio, and M.~Merimaa.
\newblock Frequency-comb-referenced tunable diode laser spectroscopy and laser
  stabilization applied to laser cooling.
\newblock {\em Appl. Opt.}, 53(31):7476--7482, 2014.

\bibitem{Francis2014}
S.~P. Francis, T.~T.-Y. Lam, K.~M. A.~J. Sutton, R.~L. Ward, D.~E. McClelland,
  and D.~Shaddock.
\newblock Weak-light phase tracking with a low cycle slip rate.
\newblock {\em Opt. Lett.}, 39(18):5251--5254, Sep 2014.

\bibitem{Gerberding2013}
O.~Gerberding, B.~Sheard, I.~Bykov, J.~Kullmann, J.~J.~E. Delgado, K.~Danzmann,
  and G.~Heinzel.
\newblock Phasemeter core for intersatellite laser heterodyne interferometry:
  modelling, simulations and experiments.
\newblock {\em Classical Quant. Grav.}, 30(23):235029, 2013.

\bibitem{Hall2006}
J.~L. Hall.
\newblock Nobel lecture: Defining and measuring optical frequencies.
\newblock {\em Rev. Mod. Phys.}, 78(4):1279, 2006.

\bibitem{Hansch2006}
T.~W. H{\"a}nsch.
\newblock Nobel lecture: passion for precision.
\newblock {\em Rev. Mod. Phys.}, 78(4):1297, 2006.

\bibitem{Holzwarth2000}
R.~Holzwarth, T.~Udem, T.~W. H{\"a}nsch, J.~Knight, W.~Wadsworth, and P.~S.~J.
  Russell.
\newblock Optical frequency synthesizer for precision spectroscopy.
\newblock {\em Phys. Rev. Lett.}, 85(11):2264, 2000.

\bibitem{Hong2016}
F.-L. Hong.
\newblock Optical frequency standards for time and length applications.
\newblock {\em Meas. Sci. Technol.}, 28(1):012002, 2016.

\bibitem{Jones2000}
D.~J. Jones, S.~A. Diddams, J.~K. Ranka, A.~Stentz, R.~S. Windeler, J.~L. Hall,
  and S.~T. Cundiff.
\newblock Carrier-envelope phase control of femtosecond mode-locked lasers and
  direct optical frequency synthesis.
\newblock {\em Science}, 288(5466):635--639, 2000.

\bibitem{Katori2003}
H.~Katori, M.~Takamoto, V.~G. Pal'chikov, and V.~D. Ovsiannikov.
\newblock Ultrastable optical clock with neutral atoms in an engineered light
  shift trap.
\newblock {\em Phys. Rev. Lett.}, 91:173005, Oct 2003.

\bibitem{Kobtsev2007}
S.~Kobtsev, S.~Kandrushin, and A.~Potekhin.
\newblock Long-term frequency stabilization of a continuous-wave tunable laser
  with the help of a precision wavelengthmeter.
\newblock {\em Appl. Opt.}, 46(23):5840--5843, Aug 2007.

\bibitem{Ludlow2015}
A.~D. Ludlow, M.~M. Boyd, J.~Ye, E.~Peik, and P.~O. Schmidt.
\newblock Optical atomic clocks.
\newblock {\em Rev. Mod. Phys.}, 87(2):637, 2015.

\bibitem{Luo2015}
Y.~Luo, H.~Li, H.-C. Yeh, and J.~Luo.
\newblock A self-analyzing double-loop digital controller in laser frequency
  stabilization for inter-satellite laser ranging.
\newblock {\em Rev. Sci. Instrum.}, 86(4):044501, 2015.

\bibitem{Mcferran2018}
J.~McFerran.
\newblock Laser stabilization with a frequency-to-voltage chip for narrow-line
  laser cooling.
\newblock {\em Opt. Lett.}, 43(7):1475--1478, 2018.

\bibitem{Mcgrew2018}
W.~McGrew, X.~Zhang, R.~Fasano, S.~Sch{\"a}ffer, K.~Beloy, D.~Nicolodi,
  R.~Brown, N.~Hinkley, G.~Milani, M.~Schioppo, et~al.
\newblock Atomic clock performance enabling geodesy below the centimetre level.
\newblock {\em Nature}, 564(7734):87, 2018.

\bibitem{Oelker2019}
E.~Oelker, R.~Hutson, C.~Kennedy, L.~Sonderhouse, T.~Bothwell, A.~Goban,
  D.~Kedar, C.~Sanner, J.~Robinson, G.~Marti, et~al.
\newblock Demonstration of 4.8$\times$$10^{-17}$ stability at 1 s for two
  independent optical clocks.
\newblock {\em Nature Photon.}, 13(10):714--719, 2019.

\bibitem{Prehn2017}
A.~Prehn, R.~Glöckner, G.~Rempe, and M.~Zeppenfeld.
\newblock Fast, precise, and widely tunable frequency control of an optical
  parametric oscillator referenced to a frequency comb.
\newblock {\em Rev. Sci. Instrum.}, 88(3):033101, 2017.

\bibitem{Saleh2015}
K.~Saleh, J.~Millo, A.~Didier, Y.~Kersal\'{e}, and C.~Lacro\^{u}te.
\newblock Frequency stability of a wavelength meter and applications to laser
  frequency stabilization.
\newblock {\em Appl. Opt.}, 54(32):9446--9449, Nov 2015.

\bibitem{Tyumenev2016}
R.~Tyumenev, M.~Favier, S.~Bilicki, E.~Bookjans, R.~L. Targat, J.~Lodewyck,
  D.~Nicolodi, Y.~L. Coq, M.~Abgrall, J.~Guéna, L.~D. Sarlo, and S.~Bize.
\newblock Comparing a mercury optical lattice clock with microwave and optical
  frequency standards.
\newblock {\em New J. Phys.}, 18(11):113002, 2016.

\end{thebibliography}

\bibliographystyle{abbrv}

\end{document}